\documentclass[aps,prd,twocolumn,showpacs,floatfix]{revtex4-1}
\usepackage{graphicx}

\begin{document}

\title{Relativistic self-gravitating gas in dynamical equilibrium}

\author{Alexander B. Kashuba}

\affiliation{}

\date{\today}

\begin{abstract}
Static spherically symmetric solution of the Einstein's equations is found representing averaged properties of an infinite self-gravitating gas in the dynamical equilibrium. It depends upon three parameters: the core radius, the relativistic factor $0<z<1$, defining the density and the mass, and one structural parameter. In the relativistic limit, $z\to 1$, an open nucleus of focused gravitational fields develops with the size being a small fraction of the core radius whereas the outskirts of the self-gravitating gas remains non-relativistic. Inside the nucleus the self-gravitating gas is described by a universal perfect fluid with the relativistic one-dimensional equation of state. At $z=1$, the space and time develops a point like singularity at the center. A characteristic property of the nucleus is incompressibility. New particles added to the self-gravitating gas bounce back into the outskirts rather than fall into the center.
\end{abstract}

\pacs{} 

\maketitle

Properties of the self-gravitating gas depend fundamentally on its size and mass \cite{rufini}. Small and light systems are shaped by the electroweak and strong interactions and are reach in diversity, planets and clouds, stars and pulsars, whereas large and heavy systems are `ideal' in this sense and are more universally defined by the gravitation. All large and heavy systems are thought to end up in the process of a gravitational collapse in the state of a black hole \cite{OS,MTW}. A well studied example is a dust falling into the center \cite{Tolman}. Yet, observable large and heavy self-gravitating systems, galaxies, are different - they are sustained by the internal Jeans pressure of chaotic motion of stars in the so-called virial dynamical equilibrium \cite{bt}. If a mass is added to a non-relativistic galaxy it will speed up the internal motion of stars and accommodate to a new old state of the dynamical equilibrium. Until the internal conditions become relativistic ones. Will such relativistic self-gravitating gas collapse into a black hole at the center if burdened further by an addition of mass? Presumably, the answer should be found in the Einstein's General Relativity alone \cite{MTW}. Recently, evidence emerges \cite{ancient} that ancient galaxies of comparable mass to the contemporary galaxies are compressed by as much as ten times while in the process of evolution they have expanded rather than collapsed. This paper studies the relativistic  non-rotating self-gravitating gas.

There are divergent views in literature on the structure of galaxies. One view is that galaxies are collisionless and preserve faithfully the initial conditions of motion of constituent stars and, therefore, there are many possible structures \cite{bt,structure}. An alternative view is that there exists a hidden law governing galaxies, initial conditions are irrelevant in the longer run and all galaxies are similar \cite{simple}. Self-gravitating gas is known to equilibrate virially \cite{bt}. In Ref.\cite{hazy} this global property has been augmented by a local equipartition of the virial among subsystems, like that in the statistical mechanics, severely restricting possible structures of galaxies. For instance, in central cores of galaxies where the stellar velocity dispersion is isotropic a unique structure, the Plummer model, is only possible. Exploring the vicinity of the Plummer model in the `model space' all structures of the infinite non-relativistic self-gravitating gas has been found \cite{hazy}. Unlike that, the finite mass structures of Ref.\cite{structure} violate the equipartition of the virial explicitly. It is interesting to extend the structural studies of galaxies into the relativistic case. An ancient distant galaxy with the Doppler spectral dispersion of stellar motion of one few tens of the velocity of light has been reported \cite{relativistic}.

Classical infinite self-gravitating gas has the following particle distribution function in the phase space $(\vec{r},\vec{v})$ \cite{hazy}:
\begin{eqnarray}
f(\vec{r},\vec{v})=\frac{1}{6\pi^2 (1+4\Theta)} \frac{1}{\sqrt{2\Phi(\vec{r})-\vec{v}^2- \frac{1+4\Theta}{12\Theta} [\vec{r}\times\vec{v}]^2}} + \nonumber\\ + \frac{1}{3\pi\sqrt{2\pi}} \frac{1+6\Theta}{1+4\Theta}  \frac{\Gamma\left(6+\frac{1}{\Theta}\right)}{\Gamma\left(\frac{9}{2}+\frac{1}{\Theta}\right)}  \left(\Phi(\vec{r})- \frac{\vec{v}^2}{2}\right)^{\frac{7}{2}+ \frac{1}{\Theta}}.
\label{PDF}
\end{eqnarray}
This gas consists of an infinite number of particles moving predominantly in the averaged gravitational potential $\Phi(\vec{r})$. Close encounter between particles, collisions, correlate the gas into the state of the dynamical equilibrium. The structure of the self-gravitating gas is defined by the structural parameter $\Theta$. Since the inertial mass equals to the weight, the distribution of mass among particles here can be arbitrary, for instance, all particles having a unit mass. In the infinite self-gravitating gas the single normalizing distance is the radius of the core $r_c$, set to one in this paper. Since the particle distribution function, Eq.(\ref{PDF}), depends only on the two integrals of particle motion, the energy and the angular momentum, it is conserved and satisfies the Liouville's equation:
\begin{equation}
\frac{\partial}{\partial t}f+\vec{v}\cdot \frac{\partial}{\partial \vec{r}}f+ \frac{d\Phi}{d\vec{r}} \cdot \frac{\partial}{\partial \vec{v}}f =0.
\label{Liouville}
\end{equation}
The averaged over random motions of constituent particles gravitational potential, or rather that of the inverted sign, is:
\begin{equation}
\Phi(\vec{r})=\left(1+\frac{1+4\Theta}{12\Theta} \vec{r}^2\right)^{-\frac{2\Theta}{1+4\Theta}}.
\label{PhiProfile}
\end{equation}
It is normalized to the averaged square of random velocities. The averaged spatial density of the self-gravitating gas is found from the particle distribution function Eq.(\ref{PDF}) by integrating over all velocities:
\begin{equation}
\rho(\vec{r})=\frac{1}{4\pi} \left(1+\frac{\vec{r}^2}{36\Theta}\right) \left(1+\frac{1+4\Theta}{12\Theta} \vec{r}^2\right)^{-2-\frac{2\Theta}{1+4\Theta}}.
\label{DensityProfile}
\end{equation}
The central density here is set to one. However, in the relativistic treatment below the central density, $z/2$, will become a gauge of relativistic effects in the self-gravitating gas. The potential and the density in Eqs.(\ref{PhiProfile},\ref{DensityProfile}) are related by the Poisson's equation. In this paper the Newton gravitation constant $G$ and the velocity of light $c$ are both set to unity. 

In a relativistic spherically symmetric self-gravitating gas one defining parameter: $0<z<1$, where the limit $z=0$ is the classical one and the limit $z=1$ is the relativistic one, can be chosen to be a central redshift: $\omega_\infty=\omega_c \sqrt{1-z}$. Indeed, due to the spherical symmetry, a probe in the center is static and experiences no average force. A light emitted from the probe towards a remote observer can define conditions prevailing at the center. While particles can approach the velocity of light at the center, far away in the outskirts they move with the non-relativistic velocities. One natural conjecture to connect these two limits is that a remote observer will observe the retarded motion according to Eq.(\ref{PDF}) all the way towards the center whereas an internal observer will observe accelerated relativistic motions of particles due to the deformation of the space and time. Then, the time component of the metric tensor:
\begin{equation}
g_{00}=1-z\Phi(r),
\end{equation}
where the potential is given by Eq.(\ref{PhiProfile}), follows. Indeed, consider a circular trajectory of radius $r$ around the center in the averaged gravitational field observed from the infinity face on. The circular velocity, observed remotely as $v=r d\varphi/dt$, is known in the General Relativity to be $2 v^2=r dg_{00}/dr$ \cite{MTW}. On the other hand it equals to $v^2=-r d\Phi/dr$ in the classical mechanics.

The solution of the Einstein's equations:
\begin{equation}
R^i_j-\frac{1}{2}\delta^i_j R=\frac{8\pi G}{c^4} T^i_j
\end{equation}
is searched for in terms of a static spherically symmetric diagonal metric tensor:
\begin{equation}
g_{\theta\theta}=-r^2, \quad g_{\varphi\varphi}=-r^2 \sin^2\theta.
\label{sphere}
\end{equation}
In terms of the short-hand notations:
\begin{equation}
\mathcal{Y}(r)=\frac{12\Theta+(1+2\Theta)r^2}{12\Theta+(1+4\Theta)r^2}\ z\Phi(r),
\end{equation}
and a radially varied power index:
\begin{equation}
\Delta(r)=\frac{12\Theta+r^2}{12\Theta+(1+2\Theta)r^2}
\end{equation}
the radial component of the metric tensor is found to be:
\begin{equation}
g_{rr}=-\left(1- \mathcal{Y}(r)\right)^{\Delta(r)}/ \left(1-z\Phi(r)\right).
\end{equation}
The energy-momentum tensor of the matter, particles of the self-gravitating gas, is diagonal and follows from the Einstein's equations. The time component is related to the radial component of the space and time metric tensor:
\begin{equation}
\frac{8\pi G}{c^4}\frac{1}{r}\int_0^r\ T^0_0(r)\ r^2\ dr=1+\frac{1}{g_{rr}(r)}.
\label{Tolman}
\end{equation}
The radial component of the matter energy-momentum tensor is found as:
\begin{equation}
T^r_r(r)=\frac{1}{8\pi r^2}\left(1-\frac{1-\frac{12\Theta+r^2}{12\Theta+(1+4\Theta)r^2}\ z\Phi(r)}{(1- \mathcal{Y}(r))^{\Delta(r)}}\right),
\end{equation}
whereas the two tangential components are equal and are found as:
\begin{equation}
T^\varphi_\varphi(r)=-\frac{1}{48\pi}\frac{\mathcal{Y}(r)+\left(1-\mathcal{Y}(r)\right)\log\left(1-\mathcal{Y}(r)\right)}{(1- \mathcal{Y}(r))^{\Delta(r)}\ \left(1+\frac{1+2\Theta}{12\Theta}r^2\right)^2}.
\end{equation}
The particle distribution function Eq.(\ref{PDF}) represents the boundary conditions at $r\to\infty$ for the above solution of the Einstein's equations, which account in full details for the relativistic dynamics of the self-gravitating gas.

On the outskirts, at $r\to \infty$ and $\Phi(r)\to 0$, there persists a hallo of the self-gravitating gas and the space and time slowly approaches the flat Galilean ones:
\begin{equation}
g_{rr}\approx -\left(1-\frac{1+2\Theta}{1+4\Theta} z\Phi(r)\right)^{\frac{1}{1+2\Theta}}/\left(1-z\Phi(r)\right).
\end{equation}
Therefore, the total Arnowitt-Deser-Misner mass encircled by a large radius for a relativistic self-gravitating gas at $z\to 1$ is given by the integral of the non-relativistic density Eq.(\ref{DensityProfile}) over the flat space and setting a proper parameter $z$ in the end. A remote observer observes the spatial distribution of particles according to the Eq.(\ref{DensityProfile}), depending on the structural parameter $\Theta$. Let us introduce a projected from the outskirts central density:
\begin{equation}
\rho_0= \frac{c^2 z}{8\pi G r_c^2}.
\label{limit}
\end{equation}
It defines the maximum of mass that the self-gravitating gas can store at the center for $z=1$ and a given core radius. If this limit is reached then the self-gravitating gas will decrease the structural parameter $\Theta$ or increase the core radius $r_c$ in order to accommodate more mass.

Novel General Relativity features of this solution are displayed in the central region. Here the spatial components of the matter energy-momentum tensor are isotropic. Describing this situation by a perfect fluid with the energy density $\epsilon$ and the pressure $p$ we find the equation of state at the center:
\begin{equation}
\frac{8\pi G}{c^4}\left(\epsilon-3p\right) r_c^2=\log\left(1+\frac{8\pi G}{c^4} \left(\epsilon+3p\right) r_c^2 \right).
\end{equation}
For large core sizes it is the three dimensional perfect fluid: $\epsilon\approx 3p$. Off the center, in the leading approximation the perfect fluid condition persists and the Tolman Oppenheimer Volkoff equation holds \cite{MTW}. Initially, the space and time metric tensor evolves as:
\begin{equation}
g_{00}\approx 1-z+\frac{1}{6}r^2, \qquad g_{rr}\approx -\frac{1-z+r^2/3}{1-z+r^2/6}.
\end{equation}
The determinant of the space and time metric tensor develops a singularity in the relativistic limit $z=1$ exactly at the center. All components of the matter energy-momentum tensor diverges at the center:
\begin{equation}
T^r_r(r)\approx T^\varphi_\varphi(r)\approx -\frac{1}{48\pi}\frac{1}{1-z+r^2/3}.
\label{Tcenter}
\end{equation}
In a wide interval of distances from the center while still being inside the core:
\begin{equation}
\sqrt{1-z}\ r_c\ll r \ll r_c,
\label{nucleus}
\end{equation}
the self-gravitating gas can be well represented by a simple and universal, i.e. not dependent on the structural parameter $\Theta$ except for the limit $\Theta\to 0$, model: $g_{00}= r^2/6$, $g_{rr}=-2$ and the spherical metric tensor Eq.(\ref{sphere}). It is the spherically symmetric analog of the Rindler metric tensor. The scalar curvature of this space and time is positive: $R=1/r^2$. The matter is a spherically symmetric isotropic perfect fluid described by the intrinsic relativistic one-dimensional equation of state:
\begin{equation}
\epsilon=p=\frac{1}{16\pi r^2},
\end{equation}
where both the energy density and the pressure grow towards the center, Eq.(\ref{Tcenter}). The radial force applied by a thin spherical shell onto the adjacent shell is $c^4/4G$ in this model, the value conjectured in Ref.\cite{maxF} to be the maximum force in the General Relativity. The region of space Eq.(\ref{nucleus}) can be called a relativistic nucleus inside the core of the self-gravitating gas. The ratio of the spatial to the time components of the matter energy-momentum tensor reaches the maximum at
\begin{equation}
r_{nucleus}=\left(-\frac{9}{5}\frac{1-z}{\log(1-z)} \right)^{1/4} r_c
\end{equation}
distance from the center. This spherical surface can define the middle size of the relativistic nucleus.

Inside the nucleus, according to Eq.(\ref{Tolman}), a remote observer observes a growth of the ADM encircled mass:
\begin{equation}
m(r)= \frac{c^2}{4G}r,
\label{tempo}
\end{equation}
with the radial distance. The universal gradient of this mass growth is half of that for the event horizon of the Schwartzschild solution \cite{MTW} and can not be exceeded. A remote observer will observe the nucleus structure as incompressible hardcore, Eq.(\ref{tempo}). If the Bekenstein's universal upper bound \cite{Bek} is saturated for the random gas then the temperature $dE/dS$ of the self-gravitating gas inside the relativistic nucleus can be defined:
\begin{equation}
T=\frac{\hbar c}{4\pi r}.
\end{equation}
An important property of the nucleus is that it is an open system unlike the black hole. Particles of the self-gravitating gas freely enter, spend time in and leave the nucleus. Also, the number of particles should be sufficiently large for their collisions to be negligible. The core as a whole remains more or less non-relativistic as the outskirts also do. At the core radius the matter motion transforms to anisotropic along the radial direction.

In conclusion, the structure of the relativistic self-gravitating gas in the dynamical equilibrium is worked out. It reveals a tight and open nucleus described by the relativistic one-dimensional equation of state, an alternative to the black hole. This nucleus is incompressible to a remote observer. There are no singularities of the space and time and the relativistic self-gravitating gas in the dynamical equilibrium can grow to any sizes and masses without undergoing the gravitational collapse.


\begin{thebibliography}{99}

\bibitem{rufini} R. Ruffini and S. Bonazzola, Phys. Rev., \textbf{187}, 1767 (1969)

\bibitem{OS} J. R. Oppenheimer and H. Snyder, Phys. Rev. \textbf{56}, 455 (1939)

\bibitem{MTW} Ch. W. Misner,  K. S. Thorne,  J. A. Wheeler, `Gravitation', (W. H. Freeman and Co, San Francisco, 1973)

\bibitem{Tolman} R. C. Tolman, Proc. Natl. Acad. Sci., \textbf{20}, 410 (1934)

\bibitem{bt} J. Binney and S. Tremaine, `Galactic Dynamics'
(Princeton University Press, Princeton NJ, 1987)

\bibitem{ancient} B. H. C. Emonts, M. D. Lehnert, M. Villar-Martin et al.,  Science, \textbf{354}, 1128 (2016) 

\bibitem{structure} S. Tremaine. D. O. Richstone, Y.-I. Byun, A. Dressler, S. M. Faber, C. Grillmair, J. Kormendy and T. R. Lauer, Astron.J. \textbf{107}, 634 (1994) 

\bibitem{simple} M. J. Disney, J. D. Romano, D. A. Garcia-Appadoo, A. A. West, J. J. Dalcanton and L. Cortese, Nature \textbf{455}, 1082 (2008)

\bibitem{hazy} A. Kashuba, arxiv:1702.05429

\bibitem{relativistic} B. Trakhtenbrot, C. M. Urry, F. Civano, D. J. Rosario, M. Elvis, K. Schawinski, H. Suh, A. Bongiorno and B. D. Simmons, Science, \textbf{349}, 168 (2015)

\bibitem{maxF} G.W. Gibbons, Found. Phys. \textbf{32}, 1891 (2002)

\bibitem{Bek} J.D. Bekenstein, Phys.Rev.D \textbf{23}, 287 (1981)

\end{thebibliography}
\end{document}